\documentclass[a4paper,11pt]{article}

\usepackage{jcappub}
\usepackage[T1]{fontenc}
\usepackage{booktabs}
\usepackage{graphicx}
\usepackage{amsmath,amssymb}
\usepackage{float}

\title{\boldmath Evolution of first-interaction second-harmonic anisotropy in cosmic-ray air showers}

\author{Hao Sun}
\author[1]{and Cunfeng Feng\note{Corresponding author.}}
\affiliation{Institute of Frontier and Interdisciplinary Science,\\
Shandong University,\\
72 Binhai Road, Qingdao, China}

\emailAdd{sun.h@mail.sdu.edu.cn}
\emailAdd{fengcf@sdu.edu.cn}

\abstract{The longitudinal evolution of second-harmonic anisotropy inherited from the first interaction in extensive air showers is investigated with CORSIKA simulations. Events with large initial anisotropy are selected. Second-harmonic observables are evaluated separately using the coordinate and momentum azimuths at several atmospheric depths and compared with their corresponding initial harmonic orientations. More than $84\%$ of the initial second-harmonic signal is attenuated within the first $20\,\mathrm{g\,cm^{-2}}$ of shower development. In coordinate space, the second-harmonic correlation with the initial orientation persists to ground level at approximately $0.09\%$ of its initial magnitude. This residual is evaluated against a matched source-isotropized reference, constructed for each selected event by independently randomizing the azimuths of the first-interaction final-state particles and simulating the shower under identical atmospheric conditions. The positive coordinate-space ensemble-mean difference at ground level is statistically resolved, and the two samples show little event-level separation. In momentum space, a weak ensemble-level residual remains at $200\,\mathrm{g\,cm^{-2}}$, and the original and source-isotropized signals become statistically indistinguishable by $300\,\mathrm{g\,cm^{-2}}$ and remain so at ground level.}


\begin{document}
\maketitle
\flushbottom

\section{Introduction}
\label{sec:intro}

Modern extensive-air-shower (EAS) observatories, including LHAASO, IceCube/IceTop, and the Pierre Auger Observatory, now provide a broad set of measurements for individual cosmic-ray showers. They reconstruct the primary energy and mass composition and measure shower observables such as the electromagnetic and muonic particle numbers, their arrival-time and lateral distributions, and the longitudinal development of the cascade \cite{LHAASOSpectrum2024,LHAASOProton2025,IceCubeComposition2019,AugerMass2025}. These measurements inform astrophysical studies of cosmic-ray spectra, composition, and origins. They also probe the hadronic interactions that drive shower development, providing access to quantum chromodynamics (QCD) at high energies \cite{AugerHadronic2016}.

This capability motivates extending EAS studies to QCD phenomena explored in collider experiments. Relativistic heavy-ion collisions at RHIC and the LHC provide a controlled setting for studying QCD at high energy density and produce a strongly interacting quark--gluon plasma (QGP), whose collective expansion and hadronization affect final-state particle distributions \cite{Busza2018,heinz2013}. Recent oxygen--oxygen and neon--neon measurements show significant anisotropic flow in systems much lighter than Pb--Pb, supporting the relevance of collective dynamics in light-ion collisions \cite{ALICELightIon2025}. High-energy cosmic rays undergo hadron--nucleus or nucleus--nucleus collisions with atmospheric nuclei. This raises the question of whether analogous collective or dense-QCD dynamics can leave an imprint on the first interaction of an air shower \cite{lahurd2018}. If present, such effects could modify the subsequent hadronic and electromagnetic cascade and the ground-level muon yield \cite{baur2023,ManshandenStrangeball2023}.

Such a signal would extend QGP studies to a naturally occurring cosmic-ray--air collision, providing a possible route to testing QGP-like collective behavior outside accelerator-controlled systems. Cosmic rays also provide a broad and effectively continuous distribution of primary energies and a mixture of projectile species, complementing collider measurements at a finite set of controlled beam energies and collision systems \cite{LHAASOSpectrum2024,IceCubeComposition2019,AugerMass2025}. This complementarity is relevant to the muon puzzle because changes in hadronization can alter the partition of energy between electromagnetic and hadronic secondaries and thereby the ground-level muon yield \cite{AlbrechtMuonPuzzle2022,baur2023,ManshandenStrangeball2023}.

Testing this possibility requires an observable that retains information from the first interaction and leaves a detectable imprint in the shower particles. Most collider QGP observables rely on collision-level reconstruction unavailable to a ground array. A global angular correlation shared by many secondary particles is more likely to leave a measurable remnant. Azimuthal anisotropy is such an observable because it is encoded collectively in the angular distribution of produced particles. This distribution is conventionally expanded as
\begin{equation}
    \frac{\mathrm{d}N}{\mathrm{d}\phi}
    \propto
    1 + 2\sum_{n=1}^{\infty} v_n
    \cos\!\left[n\left(\phi-\Psi_n\right)\right],
    \label{eq:fourier_intro}
\end{equation}
where $v_n$ and $\Psi_n$ denote the magnitude and orientation of the $n$th harmonic. The second coefficient, $v_2$, describes the elliptic component and is a principal observable of collective behavior in relativistic nuclear collisions \cite{heinz2013}. Its event-wide character makes it a plausible carrier of first-interaction information. Shower evolution can dilute, rotate, or erase that information through successive interactions, decays, energy loss, and electromagnetic multiplication. Before $v_2$ can be used as an interaction-sensitive EAS observable, one must therefore determine how an initial second-harmonic signal develops through the shower and which component survives to experimentally relevant depths.

Several studies have taken initial steps toward this question. LaHurd and Covault compared showers initiated by EPOS-LHC events with pronounced second-harmonic structure to a reference sample and found that ground-level muon differences could persist for unusually deep first interactions. The differences weakened at higher interaction altitudes \cite{lahurd2018}. Nie et al. subsequently implemented collective-flow observables in CORSIKA simulations and evaluated $v_2$ for charged particles and muons in ultrahigh-energy showers \cite{nie2021}. A preliminary study also compared anisotropic signatures at different observation heights, showing that atmospheric depth is central to their survival \cite{sun2025}. These studies construct harmonic quantities from EAS particles and show that interaction-level modifications can affect later shower observables. The inherited component associated with a prescribed first-interaction anisotropy remains unresolved. Shower development can generate a nonzero observation-level $v_2$, so this quantity mixes newly generated structure with first-interaction memory. The present work addresses this specific question.

Separating this inherited component is nontrivial because a later harmonic may also arise from finite-multiplicity fluctuations, cascade dynamics, shower geometry, or detector effects. Isolating the contribution associated with the prescribed first-interaction anisotropy requires a matched reference in which the azimuths of the same first-interaction final-state particles are randomized and the corresponding shower is simulated under identical conditions. It also requires depth-resolved tracking of both the harmonic magnitude and its alignment with the initial orientation. Momentum-space anisotropy and coordinate-space morphology must be treated separately, since their memories can evolve on different atmospheric-depth scales. These elements determine both whether a residual harmonic remains on average and whether the original and azimuth-randomized showers remain distinguishable on an event-by-event basis.

This study uses CORSIKA simulations \cite{heck1998} of air showers initiated by first interactions enriched in large initial second-harmonic anisotropy. Second-harmonic signals are extracted at multiple observation planes corresponding to successive stages of shower development. Momentum- and coordinate-space azimuths are analyzed separately for the full recorded-particle sample and for selected electromagnetic and muonic components. The analysis quantifies attenuation of the harmonic magnitude, loss of alignment with the initial event orientation, and the evolution of the particles carrying the residual signal. The goal is to establish a quantitative connection between an interaction-level angular structure and particles recorded later in the shower. The selected sample serves as an anisotropy-enriched benchmark; QGP event identity and formation probability remain outside the analysis.

\section{Simulation framework}
\label{sec:simulation}

The simulations were organized in two successive stages. First, a sample of
${}^{56}\mathrm{Fe}+{}^{14}\mathrm{N}$ first-interaction events was generated,
from which events with large initial second-harmonic azimuthal correlations
were selected. For each selected original event, a corresponding randomized
reference was generated by independently randomizing the azimuths of its
first-interaction final particles. The original event and its randomized
reference were then simulated separately under identical air-shower
conditions. Secondary particles were recorded at several observation planes
to follow how the initial second-harmonic anisotropy is attenuated during
shower development.

\subsection{First-interaction event simulation}

A first-interaction event denotes the final-state particle ensemble produced
in the primary collision of the incoming iron nucleus with an atmospheric
nitrogen nucleus, before those particles undergo subsequent interactions in
the atmosphere. The simulations were performed with CORSIKA 7.8000, a Monte
Carlo program for simulating the interactions, decays, and propagation of
particles in extensive air showers~\cite{heck1998}. In the present work, CORSIKA was used
both to generate the first interaction and to simulate shower development from
a stored first-interaction final-state particle list. This two-stage procedure
allowed the particle azimuths of a given first-interaction event to be
randomized before the subsequent shower was simulated.

High-energy hadronic interactions were simulated with EPOS-LHC-R. Its
core--corona framework treats dense string segments as a collectively
hadronizing core, whose expansion depends on the collision geometry. Dilute
segments hadronize through string fragmentation~\cite{epos_pierog2015}. EPOS-LHC-R
retains this core--corona framework in an updated model for air-shower simulations~\cite{epos_pierog2023,epos_werner2025}. The standard EPOS-LHC-R configuration was used, retaining
its built-in collective hadronization dynamics for the first-interaction
final state.

The first interaction was generated as a fixed-target collision of a 29~PeV
${}^{56}\mathrm{Fe}$ projectile with an ${}^{14}\mathrm{N}$ target. For this
configuration, the projectile energy per nucleon is
$E_{\rm lab}^{\rm Fe}/56$, and the corresponding nucleon--nucleon
center-of-mass energy is
\begin{equation}
  \sqrt{s_{NN}} \simeq
  \sqrt{\frac{2m_N E_{\rm lab}^{\rm Fe}}{56}}
  \simeq 1~\mathrm{TeV}.
\end{equation}
The impact parameter was unrestricted. EPOS-LHC-R generated a minimum-bias
inelastic sample with internally sampled collision geometry. The main first-interaction settings
are summarized in table~\ref{tab:first_interaction_params}.

\begin{table}[H]
\centering
\caption{First-interaction simulation parameters.}
\label{tab:first_interaction_params}
\begin{tabular}{ll}
\toprule
Parameter & Value \\
\midrule
Projectile & ${}^{56}\mathrm{Fe}$ \\
Target & ${}^{14}\mathrm{N}$ \\
Projectile laboratory energy & 29~PeV \\
Nucleon--nucleon center-of-mass energy & $\sqrt{s_{NN}}\simeq 1$~TeV \\
First-interaction model & EPOS-LHC-R \\
\bottomrule
\end{tabular}
\end{table}

A total of 20,000 first-interaction events were generated. For each event,
the particle species and three-momenta of the final-state particles were
stored for event selection and the subsequent air-shower simulations.

\subsection{Selection of events}

The multiplicity $M$ was defined as the number of charged first-interaction
final-state particles entering the harmonic calculation. Events were required
to satisfy $M\geq 100$ before ranking, reducing the large finite-multiplicity
fluctuations associated with low-multiplicity events.

For each event, the transverse-momentum-weighted second-harmonic vector was
defined as
\begin{equation}
  Q_{2,w}=\sum_{i=1}^{M} w_i e^{2i\phi_i},\qquad
  W_1=\sum_{i=1}^{M} w_i,\qquad
  W_2=\sum_{i=1}^{M} w_i^2,\qquad
  w_i=p_{T,i},
\end{equation}
where $\phi_i$ is the momentum-space azimuth of particle $i$. The weighted
second-harmonic two-particle correlation was calculated as
\begin{equation}
  c_{2,w}\{2\}=\frac{|Q_{2,w}|^2-W_2}{W_1^2-W_2}.
\end{equation}
The subtraction of $W_2$ removes the contribution from pairing a particle
with itself, so $c_{2,w}\{2\}$ is a $p_T$-weighted second-harmonic
two-particle azimuthal correlation. At the ensemble level,
$\langle c_{2,w}\{2\}\rangle$ is the corresponding weighted two-particle
second-harmonic correlation. In this work, $c_{2,w}\{2\}$ ranks
first-interaction events. The $p_T$ weight enters at this initial-event
ranking stage. The chosen correlation removes the positive self-pair baseline
present in the normalized Q-vector magnitude $|Q_{2,w}|/W_1$ at finite
multiplicity.

The first-interaction final-state particle ensemble that seeds the subsequent
CORSIKA shower is termed the source distribution. The unmodified and azimuth-randomized versions
are called the original source and the source-isotropized reference,
respectively.

For the source distributions defined above, the same particles were also
characterized by the unweighted complex vector and its magnitude,
\begin{equation}
  q_2=\frac{1}{M}\sum_{i=1}^{M}e^{2i\phi_i},\qquad
  v_2=|q_2|.
  \label{eq:source_v2}
\end{equation}
This displayed $v_2$ is distinct from the $Z_{c_2}$ variable used for event
ranking.

\begin{figure}[htbp]
    \centering
    \includegraphics[width=\linewidth]{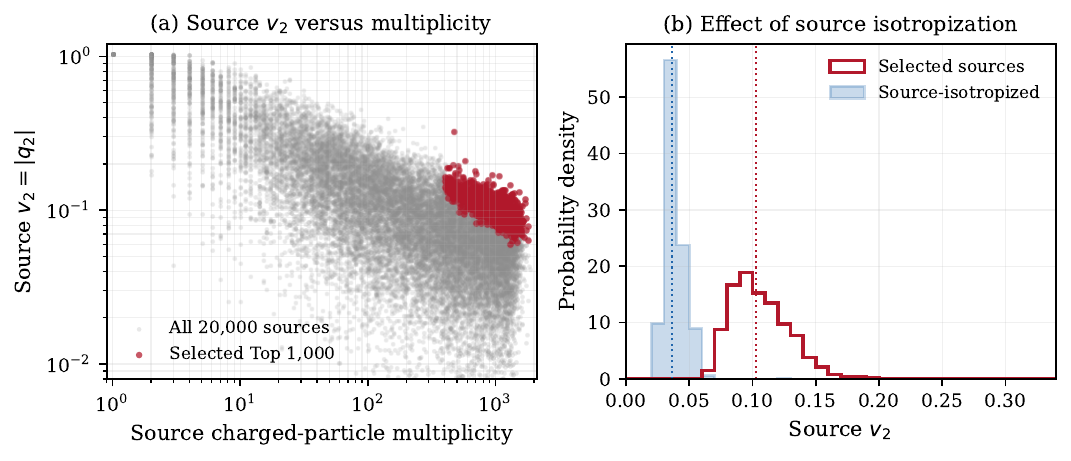}
    \caption{First-interaction source-event selection and isotropization. (a) Source $v_2$ versus charged-particle multiplicity for all generated events and the retained sample. (b) Source $v_2$ distributions for the retained events and their source-isotropized references.}
    \label{fig:source_selection}
\end{figure}

For each event, repeated azimuthal randomization provided a randomized mean
$\mu_{c_2}^{\rm rand}$ and standard deviation $\sigma_{c_2}^{\rm rand}$.
The ranking variable was defined as
\begin{equation}
  Z_{c_2}=\frac{c_{2,w}\{2\}-\mu_{c_2}^{\rm rand}}
  {\sigma_{c_2}^{\rm rand}}.
\end{equation}
This standardization accounts for event-by-event fluctuations associated with
finite multiplicity and the particle weights. Among the events satisfying
$M\geq 100$, the 1000 events with the largest $Z_{c_2}$ values were retained.
Of the 20,000 generated events, 13,613 satisfy this multiplicity requirement,
so the retained 1,000 events correspond to $7.35\%$ of the eligible sample.
The resulting sample is intentionally enriched in strong initial
second-harmonic correlations and serves as an anisotropy-enriched benchmark
with a selection distribution distinct from that of unbiased cosmic-ray events.
The selection and the resulting $v_2$ distributions are shown in
figure~\ref{fig:source_selection}.

\subsection{Source-isotropized reference events}

For each selected original event, a source-isotropized reference was generated by
assigning every first-interaction final particle a new azimuth sampled
independently from a uniform distribution on $[0,2\pi)$:
\begin{equation}
  \phi_i\longrightarrow\phi_i^{\rm rand},\qquad
  \phi_i^{\rm rand}\sim\mathcal{U}(0,2\pi).
\end{equation}
The transverse components in the source-isotropized reference were reassigned as
\begin{equation}
  p_{x,i}^{\rm rand}=p_{T,i}\cos\phi_i^{\rm rand},\qquad
  p_{y,i}^{\rm rand}=p_{T,i}\sin\phi_i^{\rm rand}.
\end{equation}
The randomization changed the transverse directions and retained the particle
identity, event multiplicity, transverse-momentum magnitudes, and longitudinal
kinematics. Azimuthal correlations are the broader class of
event-level angular correlations: collective anisotropic flow contributes to
them, together with nonflow correlations. Independent randomization removes
the original event-level azimuthal correlations as a whole, including both
components, in expectation.

Independent azimuthal draws remove the event-wise transverse-momentum
conservation and nonflow correlations present in the original final state.
The construction therefore isolates the effect of the original azimuthal
arrangement as an analysis reference. A full physical
counterfactual would require preserving these event-wise constraints during
the randomization. At each observation depth, the source-isotropized
reference provides the baseline for distinguishability. Statistical
indistinguishability of the two second-harmonic observables at a given depth
marks the loss of the retained imprint of the initial azimuthal correlations.

\subsection{Air-shower simulation}

The first-interaction final particles from the original event and its
source-isotropized reference were supplied separately to CORSIKA through the
STACKIN option. CORSIKA initialized each shower from the supplied particle
list, with the first interaction fixed externally. Each resulting shower was
simulated with the same CORSIKA physical settings and random-seed values as
its matched partner. The two branches differed at injection in the azimuthal
directions of the supplied first-interaction particles; all other inputs were
held fixed.

The first-interaction final particles were injected at an altitude of
12,000~m above sea level, and all showers were vertical. The U.S. Standard
Atmosphere was used. The geomagnetic field was disabled to avoid the
deflection of charged particles. EPOS-LHC-R was used for subsequent
high-energy hadronic interactions. UrQMD was used for the low-energy hadronic
cascade~\cite{urqmd,urqmd1999}. The main air-shower settings and observation
levels are summarized in table~\ref{tab:air_shower_settings}.

The CORSIKA tracking thresholds were set to 0.3, 0.3, 0.003, and 0.003~GeV
for hadrons, muons, electrons, and photons, respectively. The simulations
used EGS4 for the electromagnetic cascade, with the corresponding NKG
configuration and EGSDAT6\_3 data set. Unthinned particle tracking was used
throughout. These settings were identical for the original and
source-isotropized branches.

A reference observation plane was placed 1~m below the injection point to
characterize the shower before appreciable atmospheric development.
Denoting its atmospheric depth by $X_{\rm ref}$, the shower depth used
throughout this work is
\begin{equation}
  \Delta X=X-X_{\rm ref}.
\end{equation}
The remaining observation planes correspond to nominal depth increments of
20, 40, 60, 80, 100, 200, 300, and 400~g~cm$^{-2}$. Because the showers are
vertical, $\Delta X$ is also the atmospheric-depth increment along the shower
axis.

At every observation plane, the particle species, position, energy, and
three-momentum of each secondary particle were recorded. The recorded lists
precede detector response, array acceptance, trigger, and event reconstruction.

\begin{table}[H]
\centering
\caption{Air-shower settings and observation depths.}
\label{tab:air_shower_settings}
\small
\begin{tabular}{@{}p{0.29\linewidth}p{0.64\linewidth}@{}}
\toprule
Parameter & Value \\
\midrule
Injection altitude & 12,000~m a.s.l. \\
Zenith angle & $0^\circ$ \\
Atmospheric model & U.S. Standard Atmosphere \\
Geomagnetic field & Disabled \\
\midrule
$\Delta X$ (g~cm$^{-2}$) & 0, 20, 40, 60, 80, 100, 200, 300, 400 \\
Altitude (m a.s.l.) & 11,999, 11,387, 10,828, 10,315, 9,836, 9,378, 7,390, 5,770, 4,400 \\
\bottomrule
\end{tabular}
\end{table}

\section{Harmonic observables and statistical methodology}
\label{sec:observables}

At each observation depth, second-harmonic observables were calculated in
momentum space and coordinate space to quantify both the azimuthal-anisotropy
magnitude and the part that remains correlated with the first-interaction
orientation. For a particle recorded at an observation plane, the two azimuthal
angles are
\begin{equation}
  \phi_p=\operatorname{atan2}(p_y,p_x),\qquad
  \phi_r=\operatorname{atan2}(y,x),
\end{equation}
where $(p_x,p_y)$ are its transverse-momentum components and $(x,y)$ are its
coordinates relative to the shower axis. The observables were evaluated for
the full recorded-particle sample and separately for two selected components,
electromagnetic particles $(\gamma,e^{\pm})$ and muons $(\mu^{\pm})$. The principal analysis used the
complete recorded-particle sample. Energy and transverse-momentum intervals
were reserved for the differential study.

At each observation plane, $q_2$ and $v_2$ were calculated with the definition
in eq.~\eqref{eq:source_v2}, using either $\phi=\phi_p$ or $\phi=\phi_r$ and
unit particle weights. The corresponding second-harmonic orientation is
$\Psi_2=\tfrac{1}{2}\arg q_2$. This is the standard Q-vector representation
of azimuthal harmonics~\cite{poskanzer1998,bilandzic2011}.
The complex vector $q_2$ is retained because it contains both the magnitude
$v_2$ and the orientation $\Psi_2$; here $v_2$ denotes the event-level
normalized Q-vector magnitude. All particles have unit weight in
the observation-plane analysis.

For statistically independent particles with uniformly distributed azimuths,
the expected squared magnitude of the normalized Q-vector is exactly
\begin{equation}
  \left\langle |q_2|^2\right\rangle_{\rm rand}=\frac{1}{M}.
\end{equation}
The random-walk scale of $v_2$ therefore decreases approximately as $M^{-1/2}$~\cite{poskanzer1998,bilandzic2011}. Because the particle multiplicity changes strongly with
observation depth and particle component, raw values of $v_2$ at different
planes contain different finite-multiplicity baselines.

For event $i$, the reference vector is obtained from all recorded particles of
the original event at the reference plane $X_{\rm ref}$,
\begin{equation}
  q_{2,i}^{\rm ref}=q_{2,i}^{O}(X_{\rm ref}),\qquad
  \Psi_{2,i}^{\rm ref}=\frac{1}{2}\arg q_{2,i}^{\rm ref}.
\end{equation}
It is constructed independently in momentum space and coordinate space. The
label $O$ denotes the original shower. The label $I$ denotes the shower
propagated from the source-isotropized reference defined in the Simulation
section and is referred to below as the source-isotropized shower. In each azimuth space, both showers are
evaluated relative to the same first-interaction reference vector at every
observation depth.

For branch $a$, the aligned amplitude of event $i$ is first defined as
\begin{equation}
  A_i^{a}(X)=\operatorname{Re}\!\left[
  q_{2,i}^{a}(X)e^{-2i\Psi_{2,i}^{\rm ref}}\right],
  \qquad a\in\{O,I\}.
\end{equation}
Here $A$ denotes the signed event-level amplitude aligned with the original
first-interaction orientation. It has the same units as the normalized Q-vector;
its normalization is independent of the reference magnitude.

The corresponding ensemble retention of the initial complex harmonic is
\begin{equation}
  H_2^{a}(X)=
  \frac{\displaystyle\sum_i q_{2,i}^{a}(X)
        q_{2,i}^{\rm ref*}}
       {\displaystyle\sum_i |q_{2,i}^{\rm ref}|^2},
  \qquad a\in\{O,I\}.
\end{equation}
Because $q_{2,i}^{\rm ref}=|q_{2,i}^{\rm ref}|
e^{2i\Psi_{2,i}^{\rm ref}}$, its real part is the reference-amplitude-weighted,
reference-power-normalized ensemble sum of the aligned amplitudes,
\begin{equation}
  \operatorname{Re}H_2^{a}(X)=
  \frac{\displaystyle\sum_i |q_{2,i}^{\rm ref}|A_i^{a}(X)}
       {\displaystyle\sum_i |q_{2,i}^{\rm ref}|^2}.
\end{equation}
The quantities $A_i^{a}$ and $H_2^{a}$ describe the same reference-axis
projection at different statistical levels. $A_i^{a}$ is a real event-level
quantity. $H_2^{a}$ is a normalized complex ensemble quantity. This
cross-product construction follows the logic of scalar-product Q-vector
correlations~\cite{luzum2013}, with a normalization chosen so that
$H_2^{O}(X_{\rm ref})=1$. Its real part decreases when the harmonic magnitude
is attenuated, when its orientation decorrelates, or through both effects; its
imaginary part measures a coherent rotation relative to the reference
orientation. The ratio-of-sums form also avoids the instability and excessive
weight that event-by-event ratios $q_2(X)/q_2^{\rm ref}$ would assign to events
with small $|q_2^{\rm ref}|$.

Azimuthal randomization is also applied to the secondary particles already recorded at
each observation plane. For every event, depth, particle component, and
azimuth space, each analyzed angle is independently redrawn from a uniform
distribution on $[0,2\pi)$. The multiplicity and remaining particle
information are retained. Three realizations are generated and averaged.
This observation-plane operation uses the same angular randomization as the
first-interaction source-isotropized reference. It acts after shower propagation
and serves a separate purpose. The first-interaction source-isotropized reference is propagated
through CORSIKA and tests whether the original and source-isotropized showers remain
distinguishable. Observation-plane randomization is performed after shower
propagation and determines the finite-multiplicity random-walk baseline of the
observable at each plane, allowing the anisotropy signal to be compared across
observation planes of the same shower.

The distinction between first-interaction randomization and observation-plane
randomization is essential when comparing different depths. A small-$M$
sample can have a comparatively large $v_2$ from random sampling alone. The
corresponding magnitude baseline is smaller for a large-$M$ sample. For the
signed reference-axis projection, the randomized expectation
is zero. The aligned amplitude of each branch is corrected event by event as
\begin{equation}
  C_i^{a}(X)=A_i^{a}(X)
  -\overline{A_i^{a,\rm rand}(X)},
  \qquad a\in\{O,I\},
\end{equation}
where the overline denotes the mean of the three observation-plane
randomizations for that event. The paired event-level difference is then
defined separately as
\begin{equation}
  D_i(X)=C_i^{O}(X)-C_i^{I}(X).
\end{equation}
The ensemble retained harmonic difference is the corresponding
reference-amplitude-weighted aggregation,
\begin{equation}
  \Delta\operatorname{Re}H_2(X)=
  \frac{\displaystyle\sum_i |q_{2,i}^{\rm ref}|D_i(X)}
       {\displaystyle\sum_i |q_{2,i}^{\rm ref}|^2}.
\end{equation}
A positive value means that the original showers retain more second-harmonic
structure aligned with their reference orientations than the matched
source-isotropized showers after the observation-plane baselines are removed.

Statistical intervals were obtained by bootstrap resampling of the
first-interaction events 10,000 times~\cite{efron1979}. For each event, the three
observation-plane randomizations were averaged first; the resulting paired
original and source-isotropized records at all depths were then resampled together.
The 2.5th and 97.5th percentiles of the resulting distributions define pointwise
95\% confidence intervals.

An ensemble-mean residual leaves the separability of the two members of an
individual shower pair unresolved. The reported event-level summaries are the
fraction of events with $D_i>0$
and the standardized mean original--source-isotropized difference
\begin{equation}
  f_{O>I}=\frac{1}{N}\sum_{i=1}^{N}\mathbf{1}(D_i>0),
  \qquad
  d_{O-I}=\frac{\overline{D}}{s_D},
\end{equation}
where $\overline{D}$ and $s_D$ are the sample mean and sample standard
deviation of the event-level differences. The quantity $f_{O>I}$ is the fraction
of paired events in which the original shower has the larger aligned signal.
The quantity $d_{O-I}$ is an effect-size measure that expresses the mean
difference in units of its event-to-event standard deviation. It describes
separation relative to event-to-event variation as a descriptive effect size.

Finally, the same observables were also calculated in fixed energy and
transverse momentum intervals to identify which secondary particles contribute
to the retained harmonic structure. The energy intervals were
\begin{equation}
  E\in[0,0.1),\ [0.1,1),\ [1,10),\ [10,\infty)~\mathrm{GeV},
\end{equation}
and the transverse momentum intervals were
\begin{equation}
  p_T\in[0,0.05),\ [0.05,0.2),\ [0.2,0.8),\ [0.8,\infty)~\mathrm{GeV}.
\end{equation}
The same intervals were used at every observation depth, and all preceding
definitions were applied independently within each interval. For the additive
bin contributions shown in the Results, the Q-vector formed in a given
interval was normalized by the inclusive multiplicity, so that the interval
contributions sum to the inclusive vector.

\section{Results}
\label{sec:results}

\subsection{Longitudinal attenuation of the retained second harmonic}
\label{sec:results_longitudinal}

The longitudinal evolution of the second-harmonic azimuthal anisotropy is first examined separately in the original showers ($O$) and the source-isotropized reference showers ($I$). This comparison serves two purposes. It illustrates the attenuation of a strong first-interaction anisotropy during shower development and checks the randomized ensemble for spurious alignment with the first-interaction reference vector. Geometric dispersion and finite-multiplicity fluctuations should average to zero in this check. For each shower, the second-harmonic vector at each observation depth is evaluated with respect to the first-interaction reference vector $q_{2,i}^{\mathrm{ref}}$ defined in section~\ref{sec:observables}.

\begin{figure}[htbp]
    \centering
    \includegraphics[width=\linewidth]{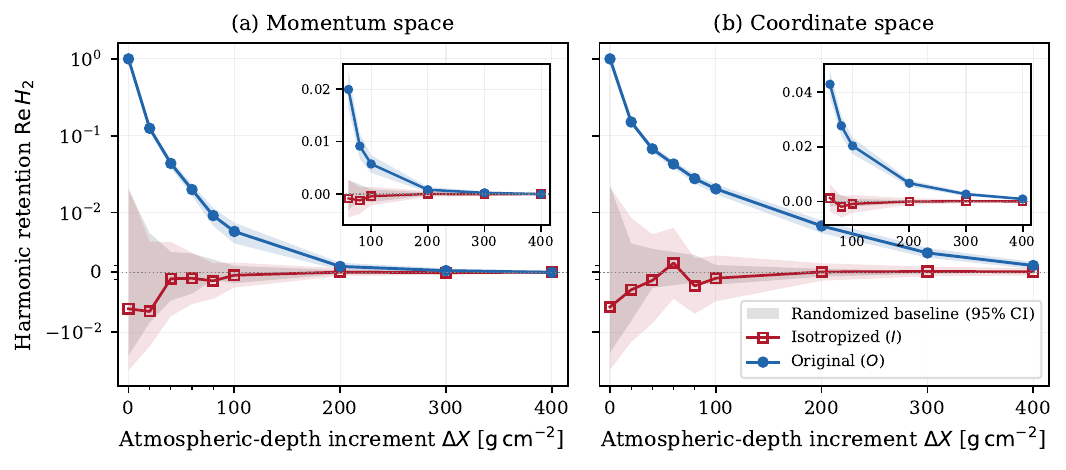}
    \caption{Longitudinal profiles of the reference-projected second-harmonic retention $\operatorname{Re} H_2$ for the inclusive particle sample defined in section~\ref{sec:observables}, shown in (a) momentum space and (b) coordinate space as a function of atmospheric-depth increment $\Delta X$. Solid circles represent the original showers ($O$) and open squares indicate the source-isotropized reference showers ($I$). Insets show the late-stage evolution ($\Delta X \ge 60~\mathrm{g\,cm^{-2}}$) on a linear scale. Shaded bands around the data denote pointwise $95\%$ bootstrap confidence intervals. The grey horizontal band represents the $95\%$ confidence region of the observation-plane randomized baseline. The horizontal dotted line marks zero.}
    \label{fig:longitudinal_evolution}
\end{figure}

The longitudinal profiles of $\operatorname{Re} H_2$ are displayed in figure~\ref{fig:longitudinal_evolution}. In the original showers, the second-harmonic structure undergoes a rapid attenuation immediately downstream of the first interaction. The attenuation is most rapid in the earliest stage of shower development: in both momentum and coordinate space, the dominant reduction occurs within the first few tens of $\mathrm{g\,cm^{-2}}$. Downstream of $\Delta X = 20~\mathrm{g\,cm^{-2}}$, the two representations exhibit distinct behaviors. The momentum-space harmonic decays steeply toward zero. The coordinate-space harmonic retains a persistent tail above the randomized baseline throughout the middle atmosphere.

The mean projection of the source-isotropized ensemble onto the corresponding original-event reference axis remains consistent with zero at every sampled depth, in agreement with the observation-plane randomized baselines. Individual source-isotropized showers can still have nonzero magnitudes because finite-multiplicity and cascade fluctuations generate event-level harmonics. The imaginary component $\operatorname{Im} H_2(X)$ is also consistent with zero across the ensemble. The double difference therefore tracks the part of the signal associated with the prescribed initial anisotropy.

\label{sec:results_survival}

The double difference $\Delta\operatorname{Re} H_2(X)$ defined in section~\ref{sec:observables} isolates the component associated with the first-interaction structure by subtracting the matched source-isotropized response and removing the observation-plane randomized baselines. The retained fraction is defined as
\begin{equation}
    R_2(X) = \frac{\Delta\operatorname{Re} H_2(X)}{\Delta\operatorname{Re} H_2(X_{\mathrm{ref}})}.
\end{equation}

\begin{figure}[htbp]
    \centering
    \includegraphics[width=\linewidth]{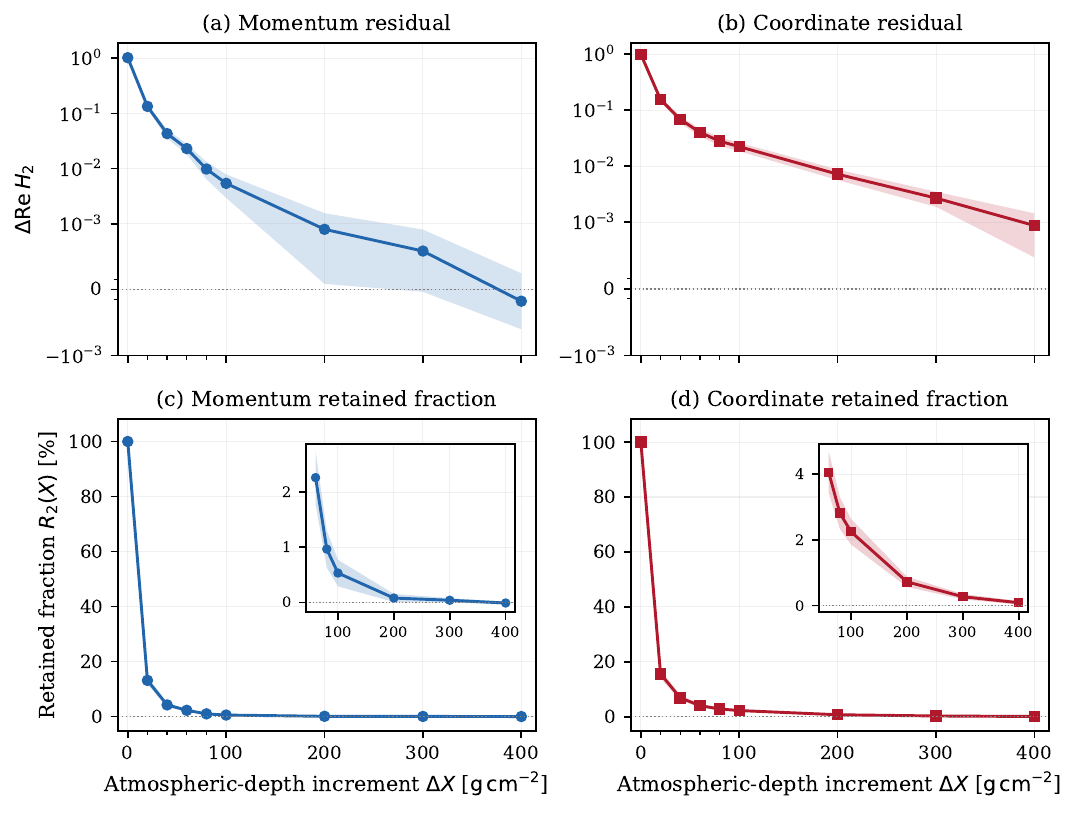}
    \caption{Longitudinal survival of the double difference $\Delta\operatorname{Re} H_2$ and retained fraction $R_2(X)$ as a function of atmospheric-depth increment $\Delta X$. Panels (a) and (b) show $\Delta\operatorname{Re} H_2$ in momentum space and coordinate space, respectively, on a symmetric logarithmic scale. Panels (c) and (d) show the corresponding retained fractions $R_2(X)$ in percent. Insets in (c) and (d) display the late-stage evolution ($\Delta X \ge 60~\mathrm{g\,cm^{-2}}$) on a linear scale. Shaded bands represent pointwise $95\%$ bootstrap confidence intervals.}
    \label{fig:contrast_survival}
\end{figure}

The longitudinal evolution of $\Delta\operatorname{Re} H_2$ and $R_2(X)$ is presented in figure~\ref{fig:contrast_survival}, and the numerical values at key atmospheric depths are summarized in table~\ref{tab:longitudinal_values}. At the reference plane ($X_{\mathrm{ref}}$, $\Delta X = 0~\mathrm{g\,cm^{-2}}$), the initial differences are
\begin{equation}
    \Delta\operatorname{Re} H_2^{\mathrm{mom}}(0) = 1.0,
    \qquad
    \Delta\operatorname{Re} H_2^{\mathrm{coord}}(0) = 0.99,
\end{equation}
both consistent with unity within statistical uncertainties.

After $\Delta X = 20~\mathrm{g\,cm^{-2}}$ of atmospheric development (corresponding to a physical distance below $620~\mathrm{m}$), the signals drop to
\begin{equation}
    \Delta\operatorname{Re} H_2^{\mathrm{mom}}(20) = 0.13,
    \qquad
    \Delta\operatorname{Re} H_2^{\mathrm{coord}}(20) = 0.15.
\end{equation}
This corresponds to retained fractions of $13\%$ in momentum space and $16\%$ in coordinate space. More than $84\%$ of the initial second-harmonic anisotropy is attenuated within the first $20~\mathrm{g\,cm^{-2}}$ of shower development. This rapid early attenuation can be understood as resulting from successive hadronic interactions, decays, and electromagnetic multiplication, which redistribute particle directions and rapidly increase the shower multiplicity.

\begin{table}[htbp]
\centering
\setlength{\tabcolsep}{5pt}
\caption{Key longitudinal values of the double difference $\Delta\operatorname{Re} H_2$ and retained fraction $R_2(X)$ with unit weight.}
\label{tab:longitudinal_values}
\begin{tabular}{r c c c c}
\toprule
$\Delta X~(\mathrm{g\,cm^{-2}})$ & $\Delta\operatorname{Re} H_2^{\mathrm{mom}}$ & $\Delta\operatorname{Re} H_2^{\mathrm{coord}}$ & $R_2^{\mathrm{mom}}$ & $R_2^{\mathrm{coord}}$ \\
\midrule
$0$   & $1.0$ & $0.99$ & $100\%$ & $100\%$ \\
$20$  & $0.13$ & $0.15$ & $13\%$  & $16\%$ \\
$40$  & $0.043$ & $0.069$ & $4.2\%$   & $6.9\%$ \\
$60$  & $0.023$ & $0.040$ & $2.3\%$   & $4.1\%$ \\
$80$  & $9.8\times 10^{-3}$  & $2.8\times 10^{-2}$  & $0.97\%$   & $2.8\%$ \\
$100$ & $5.4\times 10^{-3}$  & $2.2\times 10^{-2}$  & $0.53\%$ & $2.2\%$ \\
$200$ & $8.0\times 10^{-4}$ & $7.2\times 10^{-3}$  & $0.08\%$ & $0.72\%$ \\
$300$ & $3.9\times 10^{-4}$ & $2.7\times 10^{-3}$  & $0.04\%$ & $0.27\%$ \\
$400$ & $-1.2\times 10^{-4}$ & $8.7\times 10^{-4}$ & $-0.01\%$ & $0.09\%$ \\
\bottomrule
\end{tabular}
\end{table}

Beyond this initial stage, the momentum-space and coordinate-space signals exhibit distinct persistence scales. In momentum space, the retained difference continues its steep decline: it falls to $5.4\times 10^{-3}$ at $\Delta X = 100~\mathrm{g\,cm^{-2}}$, leaves a marginal residual of $8.0\times 10^{-4}$ at $200~\mathrm{g\,cm^{-2}}$, and becomes consistent with zero by $300~\mathrm{g\,cm^{-2}}$. At ground level, the momentum-space difference evaluates to
\begin{equation}
    \Delta\operatorname{Re} H_2^{\mathrm{mom}}(400) = -1.2\times 10^{-4},
\end{equation}
which is statistically indistinguishable from zero. Hence, by ground level, the momentum-space second-harmonic anisotropy is consistent with zero within statistical uncertainty.

In coordinate space, the initial azimuthal anisotropy persists over a substantially greater atmospheric depth. The coordinate difference remains positive and strictly larger than its momentum-space counterpart at every depth downstream of $\Delta X = 20~\mathrm{g\,cm^{-2}}$. Even at $\Delta X = 300~\mathrm{g\,cm^{-2}}$, a clearly resolved signal of $2.7\times 10^{-3}$ remains ($R_2^{\mathrm{coord}} \simeq 0.27\%$). At ground level, a finite ensemble residual survives:
\begin{equation}
    \Delta\operatorname{Re} H_2^{\mathrm{coord}}(400) = 8.7\times 10^{-4},
    \label{eq:ground_coord_val}
\end{equation}
corresponding to a retained fraction of
\begin{equation}
    R_2^{\mathrm{coord}}(400) = \frac{\Delta\operatorname{Re} H_2^{\mathrm{coord}}(400)}{\Delta\operatorname{Re} H_2^{\mathrm{coord}}(X_{\mathrm{ref}})} \simeq 0.09\%.
\end{equation}
The pointwise interval for this ensemble mean excludes zero. Its magnitude is exceptionally small, representing less than one-thousandth of the initial anisotropy at the reference plane. Its event-level implications are examined in section~\ref{sec:results_event_level}.

\subsection{Kinematic and species dependence}
\label{sec:results_differential}

The faint coordinate residual and the vanishing momentum signal motivate a breakdown by secondary-particle energy, transverse momentum, and species. Two complementary observables are evaluated: the within-bin shape observable $\Delta\operatorname{Re} H_{2,\mathrm{shape}}^{(b)}(X)$, which measures the intrinsic angular coherence among particles within kinematic interval $b$, and the additive interval contribution $C_2^{(b)}(X)$, which normalizes the harmonic vector of interval $b$ by the inclusive shower multiplicity $M$. By definition of the additive Q-vector decomposition \cite{poskanzer1998,bilandzic2011}, the interval contributions sum linearly to the inclusive difference, $\sum_b C_2^{(b)}(X) = \Delta\operatorname{Re} H_2(X)$. The two quantities separate within-bin coherence from the inclusive signal carried by each interval. Signed contributions can exceed 100\% or become negative when intervals partially cancel.

\begin{figure}[htbp]
    \centering
    \includegraphics[width=\linewidth]{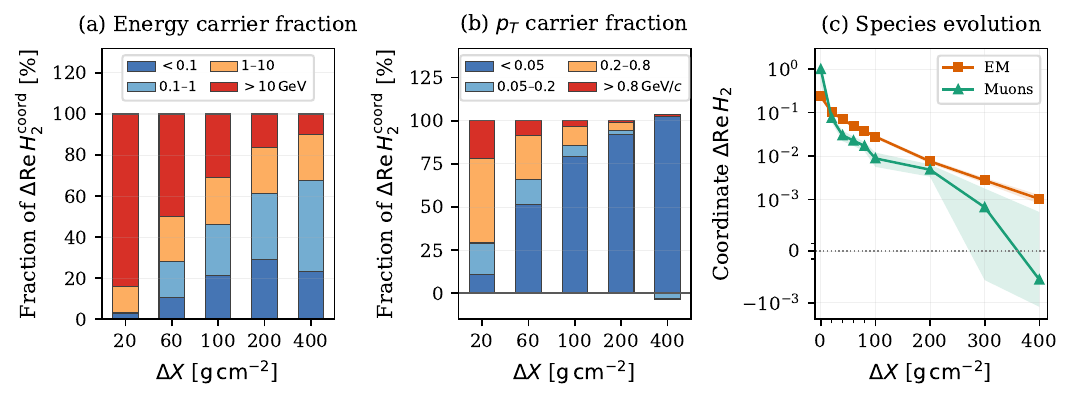}
    \caption{Kinematic and species breakdown of the coordinate-space retained difference $\Delta\operatorname{Re} H_2^{\mathrm{coord}}$. (a) Fractional energy contributions and (b) fractional transverse-momentum contributions at representative atmospheric depths. (c) Longitudinal evolution for the electromagnetic and muonic components. Shaded bands in (c) represent pointwise $95\%$ bootstrap confidence intervals.}
    \label{fig:carrier_evolution}
\end{figure}

The longitudinal evolution of the energy carriers is summarized in figure~\ref{fig:carrier_evolution}(a). In the early stage of shower development ($\Delta X = 20~\mathrm{g\,cm^{-2}}$), the retained coordinate difference is carried predominantly by energetic secondaries ($E > 10~\mathrm{GeV}$). As the cascade develops, successive electromagnetic and hadronic branchings degrade particle energies, progressively shifting the dominant share of the signal toward softer intervals. At ground level, the hierarchy is reversed: low-energy secondaries ($E < 1~\mathrm{GeV}$) dominate the surviving coordinate difference, and energetic particles make a marginal contribution. The supporting energy-dependent within-bin shape coherence curves and full differential contribution spectra across depths are presented in figures~\ref{fig:app_energy_shape} and~\ref{fig:app_differential_diagnostics} in the appendix.

The transverse-momentum distribution in figure~\ref{fig:carrier_evolution}(b) reveals a parallel transition. At early depths, the signal contains substantial contributions from moderate- and high-$p_T$ particles ($p_T > 0.2~\mathrm{GeV}/c$). Downstream, the contribution shifts systematically toward softer transverse momenta. At ground level, the coordinate residual is overwhelmingly concentrated in the lowest transverse-momentum population ($p_T < 0.05~\mathrm{GeV}/c$), which comprises the vast majority of particles in the late cascade.

Figure~\ref{fig:carrier_evolution}(c) compares two selected particle components. The electromagnetic component yields a positive coordinate difference throughout the atmosphere. The muon contribution becomes consistent with zero at ground level. These two channels support an electromagnetic origin for most of the late coordinate-space residual; other particle subdivisions are outside this comparison.

Reconciling within-bin shape coherence with inclusive additive contributions clarifies the origin of this result. High-energy particles maintain strong intrinsic angular alignment throughout the shower (as shown in appendix~\ref{sec:appendix_diagnostics}). Their late-stage multiplicity is small. Soft electromagnetic secondaries have weaker individual angular coherence. Their large multiplicity makes them the dominant carrier of the total coordinate residual. The differential breakdown shows that the dominant population shifts from energetic particles at early depth to soft, low-$p_T$ electromagnetic secondaries at ground level. The physical origin of the longer coordinate-space persistence is discussed in section~\ref{sec:discussion}.

\subsection{Event-level overlap at ground level}
\label{sec:results_event_level}

A positive coordinate difference at ensemble level leaves individual-shower separability unresolved.

To address this question, the event-level difference $D_i(X) = C_i^O(X) - C_i^I(X)$ defined in section~\ref{sec:observables} is examined across the $1,000$ simulated shower pairs, summarized by the fraction of events with positive difference $f_{O>I} = P(D_i > 0)$ and the standardized original--source-isotropized difference $d_{O-I} = \overline{D}/s_D$.

\begin{figure}[htbp]
    \centering
    \includegraphics[width=\linewidth]{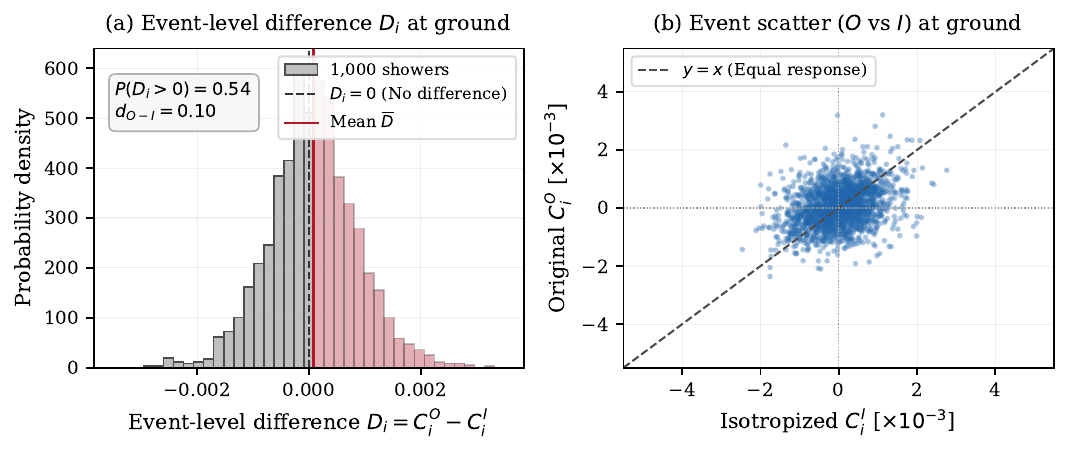}
    \caption{Event-level comparison of the coordinate-space second-harmonic observable at ground level. (a) Distribution of the paired difference $D_i = C_i^O - C_i^I$ for the $1,000$ simulated shower pairs. The dashed vertical line marks $D_i = 0$, the solid line indicates the ensemble mean $\overline{D}$, and the shaded area marks pairs with $D_i > 0$. (b) Paired values of $C_i^O$ and $C_i^I$; the dashed line marks equal response ($y = x$).}
    \label{fig:event_distinguishability}
\end{figure}

The event-level results are illustrated in figure~\ref{fig:event_distinguishability}. The distribution of event-level differences $D_i$ in figure~\ref{fig:event_distinguishability}(a) shows the fraction of shower pairs for which the original shower yields the larger coordinate harmonic aligned with the reference. This fraction is
\begin{equation}
    f_{O>I} = P(D_i > 0) = 0.54,
\end{equation}
which exceeds the equal-probability expectation of $0.50$ by $0.04$. The sample statistics evaluate to
\begin{equation}
    \overline{D} = 5.6\times 10^{-5},
    \qquad
    s_D = 5.9\times 10^{-4},
    \qquad
    d_{O-I} = \frac{\overline{D}}{s_D} = 0.095.
\end{equation}
Because the shower-to-shower geometric and cascade fluctuations ($s_D \approx 5.9\times 10^{-4}$) are an order of magnitude larger than the mean physical difference ($\overline{D} \approx 5.6\times 10^{-5}$), the distribution is overwhelmingly symmetric around zero.

The event-by-event scatter plot in figure~\ref{fig:event_distinguishability}(b) provides a direct visual confirmation of this overlap. The $1,000$ event pairs form a circular cloud centered near the origin, with individual points scattered evenly on both sides of the identity line $y = x$. The points form one overlapping population; separated or bifurcated clusters are absent.

These findings distinguish ensemble-level statistical memory from single-event observability:
\begin{enumerate}
    \item \textbf{Ensemble-level memory}: The pointwise bootstrap interval for the ensemble mean excludes zero, showing that the positive coordinate-space residual is statistically resolved for the present sample of 1,000 shower pairs.
    \item \textbf{Single-event distinguishability}: For an individual ground-level shower, event-to-event cascade fluctuations are an order of magnitude larger than the residual second-harmonic anisotropy. Discriminating an original shower from its source-isotropized reference on an event-by-event basis is therefore dominated by random fluctuations.
\end{enumerate}

In summary, a positive ensemble residual is confined to coordinate space at ground level, where its magnitude is of order $10^{-3}$ of the initial anisotropy at the reference plane ($R_2^{\mathrm{coord}} \approx 0.09\%$). The extensive event-by-event overlap shows that this residual provides little single-shower discrimination.

\section{Discussion}
\label{sec:discussion}

\subsection{Attenuation of the first-interaction anisotropy}

Figures~\ref{fig:longitudinal_evolution} and~\ref{fig:contrast_survival} show that most of the first-interaction second harmonic is erased during the earliest stage of shower development. Hadronic interactions, decays, and electromagnetic multiplication increase the particle multiplicity and redistribute particle directions. The mean projection of the source-isotropized ensemble vanishes along the original-event reference axis, indicating that cascade-generated contributions average to zero there.

The longer coordinate-space tail can be understood from the different information carried by the two observables. Momentum azimuth records the instantaneous particle direction. Transverse position integrates displacement accumulated during earlier propagation. Later interactions can therefore erase directional alignment faster than spatial morphology. A particle-ancestry analysis would be needed to test this interpretation directly.

\subsection{Particles carrying the surviving coordinate-space signal}

Figure~\ref{fig:carrier_evolution} shows that the particles carrying the residual shift from energetic secondaries at early depths toward lower energy and lower transverse momentum later in the shower. Higher-energy particles retain the larger within-bin shape signal, as shown in the appendix, and their late-stage multiplicity is small. Soft particles have weaker individual coherence. Their abundant electromagnetic population carries the ground-level residual as a weak bias.

The broad overlap in figure~\ref{fig:event_distinguishability} shows that this ensemble behavior provides little information for classifying an individual shower.

\subsection{Scope and implications}

The analysis is performed on an intentionally anisotropy-enriched benchmark. Its event distribution differs from that of a representative physical cosmic-ray sample. Source isotropization removes the complete initial azimuthal arrangement. The retained difference therefore reflects that arrangement as a whole. Collective flow is one possible component among several.

The fixed $12~\mathrm{km}$ first interaction is unusually deep and leaves less atmosphere for further shower development, favoring signal survival. Even in this configuration, a very small coordinate-space residual reaches ground. Higher first interactions are expected to attenuate it further. The vertical geometry, disabled geomagnetic field, and absence of detector response make the calculation a benchmark for intrinsic cascade attenuation. A direct ground-array prediction would additionally require these effects.

\section{Conclusion}
\label{sec:conclusion}

The simulations track a strong first-interaction second-harmonic azimuthal structure through extensive air showers and compare each original shower with a matched source-isotropized reference. Momentum-space and coordinate-space observables were analyzed separately at several atmospheric depths, allowing the part of the downstream harmonic that remains correlated with the first-interaction orientation to be quantified.

More than $84\%$ of the retained signal is lost within the first $20~\mathrm{g\,cm^{-2}}$. The momentum-space difference becomes consistent with zero before ground level. The coordinate-space signal survives longer and reaches about $0.09\%$ of its initial value at ground. The remaining coordinate-space contribution is dominated by soft, low-$p_T$ electromagnetic particles. The ensemble mean is positive. Event-level separation remains weak, with $f_{O>I}=0.54$ and $d_{O-I}=0.095$.

These results constrain the use of ground-level second-harmonic anisotropy as a probe of QGP-related dynamics in the first interaction. The simulation uses large-anisotropy events and an unusually deep $12~\mathrm{km}$ first interaction, conditions that favor signal survival. The retained difference at ground remains very small and has little event-by-event discriminating power. Any experimental interpretation would also require geomagnetic deflection and detector response.

\acknowledgments

The authors thank Huichao Song (Tsinghua University) and Zhenyu Chen (Shandong University) for helpful discussions.
This study is supported by the National Key R\&D Program of China, Ministry of Science and Technology (Grant No. 2024YFA1611401) and the National Natural Science Foundation of China (Grant No. 12675147).
OpenAI ChatGPT (GPT-5.6 Sol) was used for language polishing. The authors have carefully reviewed the AI-generated text and take full responsibility for its content.

\section*{Data and code availability}

The data and analysis code supporting the findings of this study are
available from the corresponding author upon reasonable request.

\appendix
\section{Supplementary Differential Diagnostics}
\label{sec:appendix_diagnostics}
\renewcommand{\thefigure}{A\arabic{figure}}
\renewcommand{\theHfigure}{A\arabic{figure}}
\setcounter{figure}{0}

This appendix provides supplementary diagnostic figures detailing the kinematic breakdown of the retained second-harmonic azimuthal anisotropy.

\begin{figure}[htbp]
    \centering
    \includegraphics[width=\linewidth]{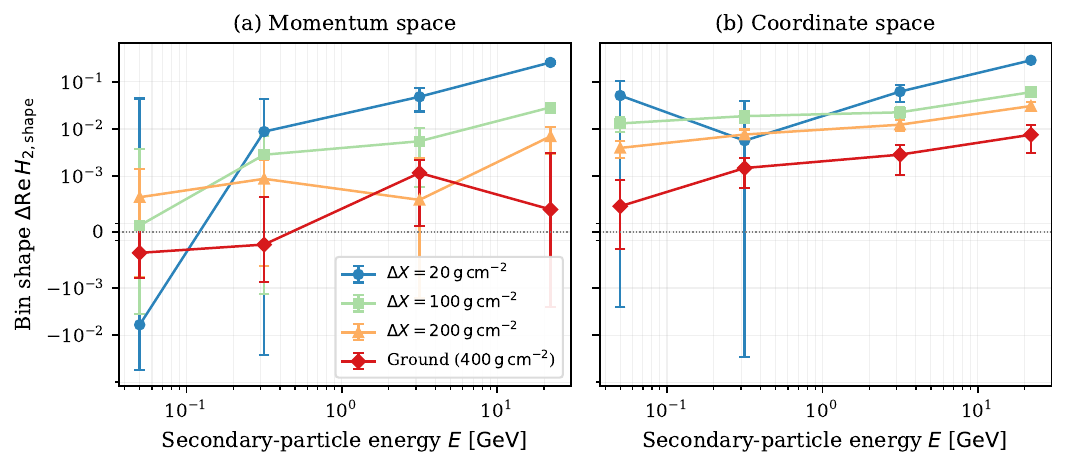}
    \caption{Secondary-particle energy dependence of within-bin second-harmonic shape coherence $\Delta\operatorname{Re} H_{2,\mathrm{shape}}^{(b)}$ in (a) momentum space and (b) coordinate space at representative atmospheric depths: $\Delta X = 20~\mathrm{g\,cm^{-2}}$ (circles), $100~\mathrm{g\,cm^{-2}}$ (squares), $200~\mathrm{g\,cm^{-2}}$ (triangles), and $400~\mathrm{g\,cm^{-2}}$ (diamonds). Error bars indicate $95\%$ bootstrap confidence intervals.}
    \label{fig:app_energy_shape}
\end{figure}

Figure~\ref{fig:app_energy_shape} displays the energy-dependent within-bin shape coherence $\Delta\operatorname{Re} H_{2,\mathrm{shape}}^{(b)}$ in momentum and coordinate space across four representative depths. The depths are 20, 100, 200, and $400~\mathrm{g\,cm^{-2}}$. In both representations, higher-energy particles generally retain stronger within-bin shape coherence with the initial first-interaction orientation.

\begin{figure}[htbp]
    \centering
    \includegraphics[width=\linewidth]{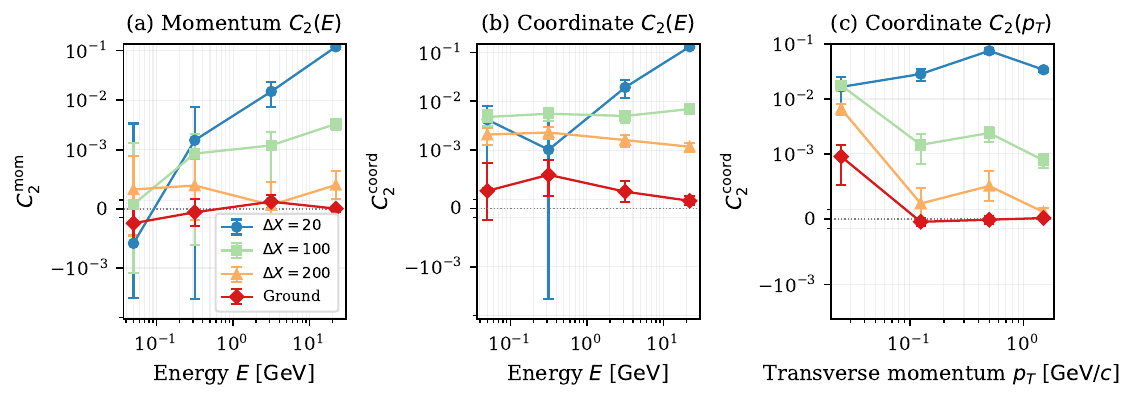}
    \caption{Additive differential contributions $C_2^{(b)}(X)$ across atmospheric depths: (a) momentum-space energy contributions $C_2^{\mathrm{mom}}(E)$, (b) coordinate-space energy contributions $C_2^{\mathrm{coord}}(E)$, and (c) coordinate-space transverse-momentum contributions $C_2^{\mathrm{coord}}(p_T)$. Error bars indicate $95\%$ bootstrap confidence intervals.}
    \label{fig:app_differential_diagnostics}
\end{figure}

\clearpage
Figure~\ref{fig:app_differential_diagnostics} displays the complete set of additive differential contribution observables $C_2^{(b)}(X)$ across energy and transverse momentum at various depths. Within-bin shape coherence remains highest for energetic particles. The additive contributions migrate toward lower energies and transverse momenta as the cascade develops, transitioning from high-$E$ dominance at early depths to low-$E$ and low-$p_T$ dominance at ground level.

\begin{thebibliography}{99}

\bibitem{LHAASOSpectrum2024}
LHAASO Collaboration,
\emph{Measurements of All-Particle Energy Spectrum and Mean Logarithmic Mass of Cosmic Rays from 0.3 to 30 PeV with LHAASO-KM2A},
\emph{Physical Review Letters} {\bf 132} (2024) 131002,
doi:10.1103/PhysRevLett.132.131002.

\bibitem{LHAASOProton2025}
LHAASO Collaboration,
\emph{Precise Measurements of the Cosmic Ray Proton Energy Spectrum in the Knee Region},
\emph{Science Bulletin} {\bf 70} (2025) 4173,
doi:10.1016/j.scib.2025.10.048,
arXiv:2505.14447.

\bibitem{IceCubeComposition2019}
IceCube Collaboration,
\emph{Cosmic Ray Spectrum and Composition from PeV to EeV Using Three Years of Data from IceTop and IceCube},
\emph{Physical Review D} {\bf 100} (2019) 082002,
doi:10.1103/PhysRevD.100.082002,
arXiv:1906.04317.

\bibitem{AugerMass2025}
Pierre Auger Collaboration,
\emph{Inference of the Mass Composition of Cosmic Rays with Energies from $10^{18.5}$ to $10^{20}$ eV Using the Pierre Auger Observatory and Deep Learning},
\emph{Physical Review Letters} {\bf 134} (2025) 021001,
doi:10.1103/PhysRevLett.134.021001,
arXiv:2406.06315.

\bibitem{AugerHadronic2016}
Pierre Auger Collaboration,
\emph{Testing Hadronic Interactions at Ultrahigh Energies with Air Showers Measured by the Pierre Auger Observatory},
\emph{Physical Review Letters} {\bf 117} (2016) 192001,
doi:10.1103/PhysRevLett.117.192001,
arXiv:1610.08509.

\bibitem{AlbrechtMuonPuzzle2022}
J. Albrecht et~al.,
\emph{The Muon Puzzle in Cosmic-Ray Induced Air Showers and Its Connection to the Large Hadron Collider},
\emph{Astrophysics and Space Science} {\bf 367} (2022) 27,
doi:10.1007/s10509-022-04054-5,
arXiv:2105.06148.

\bibitem{Busza2018}
W. Busza, K. Rajagopal, and W. van der Schee,
\emph{Heavy Ion Collisions: The Big Picture, and the Big Questions},
\emph{Annual Review of Nuclear and Particle Science} {\bf 68} (2018) 339,
doi:10.1146/annurev-nucl-101917-020852,
arXiv:1802.04801.

\bibitem{heinz2013}
U. Heinz and R. Snellings,
\emph{Collective Flow and Viscosity in Relativistic Heavy-Ion Collisions},
\emph{Annual Review of Nuclear and Particle Science} {\bf 63} (2013) 123,
doi:10.1146/annurev-nucl-102212-170540,
arXiv:1301.2826.

\bibitem{ALICELightIon2025}
ALICE Collaboration,
\emph{Evidence of Nuclear Geometry-Driven Anisotropic Flow in O--O and Ne--Ne Collisions at $\sqrt{s_{\mathrm{NN}}}=5.36$ TeV},
arXiv:2509.06428 (2025).

\bibitem{lahurd2018}
D. LaHurd and C.E. Covault,
\emph{Exploring Potential Signatures of QGP in UHECR Ground Profiles},
\emph{JCAP} {\bf 11} (2018) 007,
doi:10.1088/1475-7516/2018/11/007,
arXiv:1707.01563.

\bibitem{baur2023}
S. Baur, H. Dembinski, M. Perlin, T. Pierog, R. Ulrich, and K. Werner,
\emph{Core-Corona Effect in Hadron Collisions and Muon Production in Air Showers},
\emph{Physical Review D} {\bf 107} (2023) 094031,
doi:10.1103/PhysRevD.107.094031,
arXiv:1902.09265.

\bibitem{ManshandenStrangeball2023}
J. Manshanden, G. Sigl, and M.V. Garzelli,
\emph{Modeling Strangeness Enhancements to Resolve the Muon Excess in Cosmic Ray Extensive Air Shower Data},
\emph{JCAP} {\bf 02} (2023) 017,
doi:10.1088/1475-7516/2023/02/017,
arXiv:2208.04266.

\bibitem{nie2021}
M. Nie, H. Zhang, L. Yi, C. Feng, and Z. Xu,
\emph{Collective flow in ultra high energy cosmic rays within CORSIKA},
\emph{Proceedings of Science} {\bf ICRC2021} (2021) 459,
doi:10.22323/1.395.0459.

\bibitem{sun2025}
H. Sun and C. Feng,
\emph{Investigation of Anisotropic Flow Signatures in Extensive Air Showers at Different Observation Heights},
\emph{Proceedings of Science} {\bf ICRC2025} (2025) 1429,
doi:10.22323/1.501.1429.

\bibitem{heck1998}
D. Heck, J. Knapp, J.N. Capdevielle, G. Schatz, and T. Thouw,
\emph{CORSIKA: A Monte Carlo Code to Simulate Extensive Air Showers},
Forschungszentrum Karlsruhe Report FZKA 6019 (1998),
doi:10.5445/IR/270043064.

\bibitem{epos_pierog2015}
T. Pierog, Iu. Karpenko, J.M. Katzy, E. Yatsenko, and K. Werner,
\emph{EPOS LHC: Test of collective hadronization with data measured at the CERN Large Hadron Collider},
\emph{Physical Review C} {\bf 92} (2015) 034906,
doi:10.1103/PhysRevC.92.034906.

\bibitem{epos_pierog2023}
T. Pierog and K. Werner,
\emph{EPOS LHC-R: up-to-date hadronic model for EAS simulations},
\emph{Proceedings of Science} {\bf ICRC2023} (2023) 230.

\bibitem{epos_werner2025}
K. Werner,
\emph{EPOS.LHC-R: a global approach to solve the muon puzzle},
\emph{Proceedings of Science} {\bf ICRC2025} (2025) 358,
arXiv:2508.07105.

\bibitem{urqmd}
S.A. Bass et~al.,
\emph{Microscopic models for ultrarelativistic heavy ion collisions},
\emph{Progress in Particle and Nuclear Physics} {\bf 41} (1998) 255,
doi:10.1016/S0146-6410(98)00058-1.

\bibitem{urqmd1999}
M. Bleicher et~al.,
\emph{Relativistic hadron-hadron collisions in the Ultra-Relativistic Quantum Molecular Dynamics model},
\emph{Journal of Physics G: Nuclear and Particle Physics} {\bf 25} (1999) 1859,
doi:10.1088/0954-3899/25/9/308.

\bibitem{poskanzer1998}
A.M. Poskanzer and S.A. Voloshin,
\emph{Methods for Analyzing Anisotropic Flow in Relativistic Nuclear Collisions},
\emph{Physical Review C} {\bf 58} (1998) 1671,
doi:10.1103/PhysRevC.58.1671.

\bibitem{bilandzic2011}
A. Bilandzic, R. Snellings, and S. Voloshin,
\emph{Flow Analysis with Cumulants: Direct Calculations},
\emph{Physical Review C} {\bf 83} (2011) 044913,
doi:10.1103/PhysRevC.83.044913.

\bibitem{luzum2013}
M. Luzum and J.-Y. Ollitrault,
\emph{Eliminating Experimental Bias in Anisotropic-Flow Measurements of High-Energy Nuclear Collisions},
\emph{Physical Review C} {\bf 87} (2013) 044907,
doi:10.1103/PhysRevC.87.044907.

\bibitem{efron1979}
B. Efron,
\emph{Bootstrap Methods: Another Look at the Jackknife},
\emph{Annals of Statistics} {\bf 7} (1979) 1,
doi:10.1214/aos/1176344552.

\end{thebibliography}
\end{document}